# Empowering Web Service Search with Business Know-How

Isabelle Mirbel, Pierre Crescenzo, and Nadia Cerezo

Université de Nice Sophia-Antipolis
Laboratoire I3S (UNS/CNRS)
930 route des Colles
BP 145
F-06903 Sophia-Antipolis cedex
France

Email: {Isabelle.Mirbel,Pierre.Crescenzo,Nadia.Cerezo}@unice.fr

## Table of Contents

## Index des illustrations

## 1 Introduction

Service Oriented Computing (SOC) is a computing paradigm using and providing services to support the rapid and scalable development of distributed applications in heterogeneous environments. Despite its growing acceptance, we argue that it is difficult for business people to fully benefit of SOC if it remains at the software level. We claim it is required to move towards a description of services in business terms, i.e. intentions and strategies to achieve them and to organize their publication and combination on the basis of these descriptions.

Moreover, service providers and users still face many significant challenges introduced by the dynamism of software service environments and requirements. This requires new concepts, methods, models, and technologies along with flexible and adaptive infrastructure for services developments and management in order to facilitate the on-demand integration of services across different platforms and organizations. Business users exploit their domain expertise and rely on previous experiences to identify relevant services to fulfill new business requirements. Indeed, they develop know-how in solving software related problems (or requirements). And we claim it is required to turn this know-how into reusable guidelines or best practices and to provide means to support its capitalization, dissemination and management inside business users communities.

The ability to support adequacy between service users needs and service providers proposals is a critical factor for achieving interoperability in distributed applications in heterogeneous environments. Service final users need means to transmit their functional and non functional requirements to service designers, especially when no service is available. And service designers need means to disseminate information about available services in order to improve their acceptance by users as well as means to better handle the way business users combine services to fulfill their business goals. Reasoning about business descriptions of services and know-how about business users services combination help to support bidirectional collaboration between business users (service final users) and computer scientists (service designers).

So, from a general point of view, software engineering implies a growing need for knowledge engineering to support all aspects of software development. In this chapter, we focus on knowledge engineering to support service combination from a business user perspective.

We propose a framework, called SATIS (*Semantically AnnotaTed Intentions for Services*) [1], to capture and organize know-how about Web Services business combination. Therefore we adopt Web semantic languages and models as a unified framework to deal with business users requirements, know-how about service combination as well as Web Services descriptions. Indeed, we distinguish between intentional and operational know-how. Intentional know-how captures the different goals and sub-goals the business users try to reach during his/her combination task. Intentional know-how is specified with the help of an intentional process model [2]. Operational know-how captures the way intentional sub-goals are operationalised by available suitable Web Services. Operational know-how is formalized as queries over Web Service descriptions.

In SATIS, business users requirements, know-how about service combination as well as Web Services descriptions are resources indexed by semantic annotations [3][4][5] in order to explicit and formalize their informative content. Semantic annotations are stored into a dedicated memory. And information retrieval inside this memory relies on the formal manipulation of these annotations and is guided by ontologies.

Annotation of intentional and operational know-hows are respectively stored as abstract and concrete rules implemented as SPARQL construct queries SPARQL. When considered recursively, a set of SPARQL construct queries can be seen as a set of rules processed in backward chaining. As a result, someone looking for solutions to operationalise a business process will take advantage of all the rules and all the Web Service descriptions stored in the semantic community memory at the time of his/her search. This memory may evolve over the time and therefore the Web Services descriptions retrieved by using a rule may vary as well. Business users as well as computer scientists may both take advantage of this reasoning capability to understand the way services are combined to fulfill a business goal. This way, SATIS supports knowledge transfer from expert business users to novice ones as well as collaboration between business users and computer scientists.

Beyond an alternative way to search for Web Services, we provide means to capitalise know-how about Web Service business users combination. Another novelty of our approach is to operationalise business goals by rules in order to promote both mutualisation of specifications and cross-fertilization of know-how about Web Services business combinations. We are currently implementing our approach in the neuroscience domain where domain ontologies and semantic Web Services descriptions are already available.

In this chapter, we propose first to start by presenting a state of the art of existing approaches about scientific workflows (including neuroscience workflows) [6][7] in order to highlight business users' needs in terms of Web Services combination. Then we discuss about intentional process modeling for scientific workflows especially to search for Web Services [8][9]. Next we present our approach SATIS to provide reasoning and traceability capabilities on Web Services business combination know-how, in order to bridge the gap between workflows providers and users.

## 2   State of the Art of Scientific Workflows

Let us briefly present the current state of the art of scientific workflows research, to the best of our knowledge. We will first detail the reasons behind the creation and use of scientific workflows, then give a quick overview of the field, then present a few well-known and very active projects and finally we will highlight the challenges left to overcome.

### 2.1   Motivation

First off, computational scientific experiments obviously largely pre-date the notion of workflow. The traditional way of doing things is to gather data, adapt it to whatever program you want to run it on, run said program and post-process the outputs to analyse them. If many programs have to be chained together and the process must be repeated many times over, scripting is surely the most direct solution. It has worked for a long time, but its limitations can no longer be ignored.

One of the most obvious problems with scripting is its non-existent ease-of-use: end-users need to be fully aware of technical details of both the data they want to analyse, the programs they will use and the platforms on which the programs will run, as well as every associated transfer protocol and input/output formats. Add to this appalling list the scripting language itself and you can conclude that the end-user will spend most of his or her time dealing with computer science delicacies rather than advancing his or her own research. Another aspect of accessibility is sharing, not only of scripts but of know-how. There again scripting is as bad as it gets: either the user doesn't already know the details aforementioned and the script will be cryptic to say the least, or the user already knows them and doesn't have much to learn from the script itself.

Scientific computerized experiments, like physical experiments, have to be carried out many times, either to check the consistency of results or to try different parameters and input data or new ideas. For every modification the script author had designed the script for, all is well. Problems arise with modifications the script was not designed to cope with, because they were either dismissed or not predicted at all. Such modifications happen all the time: a program is upgraded and new protocols are adopted, new data is collected that doesn't exactly fit the previous format, new algorithms are designed that need additional parameters, etc. If the modification is small and the author is around and still has a fresh memory of his or her script, it might go well. Most often, though, it is quicker to re-write a new script than to modify an existing one. Again, more time is taken from research and lost onto computer science technicalities that are essentially not re-usable.

E-sciences of late have known a small revolution: ever increasing computational costs, dataset size and process complexity have made the traditional model of one scientist running his or her programs on a single machine obsolete. To carry out one analysis, highly-distributed heterogeneous resources are needed. Scripting obviously fails to cope with such situations. Grid technologies and SOA (Service Oriented Architecture) are partial answers to the problem but they are not well-known for their ease-of-use and further alienate the typical end-user.

Scientific workflows are gaining momentum in many data-intensive research fields, e.g.:

- Bio-informatics [10][11],
- Physics [12][13][14],
- Ecology [15],
- Disaster handling [16].

According to [17], a scientific workflow framework must fulfill eight criteria to be truly useful to the end-user, who is typically a scientist, with little knowledge in computer science, wishing to automate and share his or her scientific analyses.

| Name | Description |
|---|---|
| Clarity | Produce self-explanatory workflows |
| Well-Formedness | Help create well-formed and valid workflows |
| Predictability | Produce workflows whose behaviour is predictable |
| Recordability | Record workflow execution |
| Reportability | Keep provenance information for all results |
| Reusability | Help create new workflows from existing ones |
| Scientific Data Modeling | Provide support for scientific data modeling |
| Automatic Optimization | Hide optimization concerns from users |

While user needs vary greatly from a given field to another, we do believe those criteria to be of interest for most end-users.

### 2.2 Overview

The notion of workflow was first formally introduced in 1995 by the Workflow Management Coalition in the Workflow Reference Model [18]. It was defined as *the computerized facilitation or automation of a business process, in whole or part*. Furthermore, they explained that a *workflow is concerned with the automation of procedures where documents, information or tasks are passed between participants according to a defined set of rules to achieve, or contribute to, an overall business goal*. As is obvious from the definition itself, workflows were meant solely for business uses at the time. The concept of a scientific workflow is much more recent than the concept of workflow itself.

It is easy to mix up the notions of scientific and business workflows. Most people who know about workflows actually know about the business version which pre-dates the scientific one. A lot of core concerns and concepts are shared by both business and scientific workflows, while some issues are specific to one domain or the other [19].

Scientific Workflow systems are heavily influenced by their business counterparts, which might explain why DAG (Directed Acyclic Graphs) are the most common workflow representation in both worlds, with vertices modeling tasks and edges modeling either data or control flow. The growing need for large-scale computation, large-scale data manipulation and support for execution over highly-distributed heterogeneous resources is common to both scientific and business contexts.

In most cases, business processes are clearly defined beforehand and actual workflow building is more or less a mapping problem, whereas scientific workflow building is most often an exploratory procedure. Needs for flexibility and dynamic changes are therefore far greater for scientific workflows. While security and integrity are the top priorities in a business context, they are far outweighed by reproducibility and reusability concerns in a scientific context: a scientific experiment serves no purpose if it cannot be shared, checked (and thus, rerun) and validated by the community.

To the best of our knowledge, there have been only 4 surveys so far, detailed below, and there is no up-to-date index of existing systems.

| Date | Title | Authors | Type | #SW[1] |
|---|---|---|---|---|
| 2005 | A taxonomy of Scientific Workflow Systems for Grid Computing | J. Yu, R. Buyya,(Gridbus) | Article | 12 |
| 2006 | Scientific Workflows Survey | A. Slominski, G. von Laszewski (Java CoG Kit) | Wiki | 32 |
| 2007 | Workflows for e-science: scientific workflows for grids | I. J. Taylor et al. (Triana) | Book | 10 |
| 2008 | Scientific Workflow: A Survey and Research Directions | A. Barker, J. van Hemert | Article | 10 |

The sheer number and significant diversity of systems make it hard for a user to find the best-fit scientific workflow framework for his or her use. We believe an online comparison matrix of the most active projects would be highly beneficial to both researchers and end-users. Unfortunately, the typical lack of communication between projects makes maintenance of such a matrix a difficult task.

## 2.3  Issues

While the many scientific workflow systems share a lot, especially regarding needs and core concepts, there is no standard for either workflow, provenance or execution descriptions. Frameworks can be compared and categorized [20], but interoperability is nothing short of painful in the present state of things. This can be shown simply by the sheer variety of terms in use for the most basic element of a workflow, most often the vertices in the associated graph: operations, tasks, nodes, processes, jobs, components, services, actors, transitions, procedures, thorns, activities, elements, units [21] and probably more. It is worth noting that BPEL (Business Process Execution Language) is often cited as a strong candidate for standard workflow description language [22][23][24][25], surely because of its status as *de facto* standard business workflow description language. However, BPEL suffers many limitations and before it can be established as a standard or intermediate workflow language, it needs to be extended [25].

For the results of a scientific workflow to be of any use, they need to be reproducible and therefore fully traceable: information must be kept about the conditions in which the workflow was run, the initial input data, the intermediate results as well as the chosen parameters. This concern for traceability is commonly referred to as "provenance" and has been a research subject for a while [17][23][26][27][28][29]. Many frameworks already implement some sort of provenance tracking, but, to the best of our knowledge, a standard is yet to be established, when it is even more critical to interoperability than a common language [30]: indeed, without provenance description standards, it is near impossible to replay a given workflow execution on another system.

Portal-based access to scientific workflow technology is currently a hot topic. It indeed seems the best way to ensure accessibility to a maximum number of end-users. It might even turn out to be a necessary step for many projects, since users who are not knowledgeable about technology often shy away from installation instructions and would certainly prefer using thin-client technology [22]. Portals are sometimes already deployed [26][31][32] and often cited in Discussion and Future Works sections [22][27].

---

[1]  Number of scientific workflow systems surveyed

The problem of discovery, or how to retrieve existing workflows that can be reused or re-purposed for one's goals, cannot be solved directly inside the scientific workflow model, editor or enactor. It might matter little to the modeling expert, but for end-users who have little knowledge about workflows, it is a critical issue: it is not enough to store finished workflows in an open database with query functionalities. The surrounding platform must provide users with helpful discovery tools. Those are critical for knowledge sharing among users. Discovery of scientific workflows is a slightly underestimated research topic. To the best of our knowledge, the only team working on it is the Taverna (myGrid) team, through the myExperiment project [33][34]. Note that discovery of services, however, is a rather hot topic.

While most existing systems claim to bridge the gap between computer scientists and scientists of other fields, and huge progress has certainly been made in the ease-of-use area, the impression one gets from just looking at the most basic workflows is quite different: in most cases, the underlying Web technologies are apparent. It is hard for us to picture a user with little knowledge in computer science able to make sense of concepts such as XSLT (eXtensible Stylesheet Language Transformations). Of course, such details should be neither ignored nor hidden completely. They should be left for a lower level of abstraction and higher levels need to be created between the typical user (e.g. a neuroscientist) and the actual executable workflow, using semantic information to automate the process in the most intelligent manner possible [35].

In the two next sections, we present our approach SATIS which is a way to respond to the difficulty for end-users to discover adequate Web Services and for computer scientists to design and present their Web Services for a better visibility to users.

# 3   Intentional Process Modeling for Scientific Workflows

## 3.1   Autonomous software components for end-users

A major interest of scientific workflows is to offer an interface to combine existing autonomous software components, like Web Services. But this is not a sufficient help for end-users. We need to give us a mean to simply express their goals and to assist us to choose the pertinent components (Web Services) to implement these goals.

One of the main objectives of SATIS is to support neuroscientists when looking for Web Services to operationalise their image processing pipeline. In this section we will first discuss the role of the different actors involved in the neuroscience community and then describe the different means we provide to support neuroscientists tasks.

Three core actors are identified in our framework: the *service designer*, the *community semantic memory manager* and the *domain expert*. In a neuroscientists community, computer scientists play the roles of *service designer* and *community semantic memory manager* while neuroscientists play the role of *domain expert*.

The service designer is in charge of promoting the Web Services available in the community Web Service registry. Therefore, when s/he wants to advertise a new kind of Web Service in the neuroscientists community, in addition to adding the Web Service description in the community registry, s/he writes a generic Web Service description and associates to it high-level end-user's intentional requirements to promote the services s/he is in charge from the end-user's point of view (that is to say in a non computer scientists language). The service designer is in charge of authoring atomic reusable fragments.

The community semantic memory manager is in charge of populating the community semantic memory with reusable fragments to help domain experts to (i) specify the image processing pipelines for which they are looking for Web Services and (ii) search for Web Service descriptions to operationalise the image processing pipelines they are interested in. Indeed, s/he provides reusable fragments useful in different image processing pipelines. Basic processes, as for instance intensity corrections, common to several image analysis pipelines, are examples of such basic fragments. Therefore, s/he may look at the fragments provided by the service designer with the aim of aggregating some of them into basic image processing pipelines. For instance in the neuroscientist community, if *Image debiasing*, *Image denoising*, *Image normalisation* and *Image registration* Web Service descriptions are provided in the community Web Service registry (and associated fragments provided in the community semantic memory) at some point, the community semantic memory manager may put them together into a basic *Image preprocessing* pipeline. S/he may also identify recurrent needs when supporting domain experts in their authoring task and therefore provide adequate basic fragments for image processing pipelines.

Finally, the domain expert (or final user) is searching for Web Service descriptions to operationalise an image processing pipeline s/he is interested in. Therefore, s/he may first look in the community semantic memory if some existing fragments already deal with the main intention s/he is interested in. If another member of the community already authored an image processing pipeline achieving the same high-level goal, s/he may reuse it as is. The goal under consideration may also be covered by a larger image processing pipeline specified through a set of fragments already stored in the community semantic memory and corresponds to one of the sub-goals of the larger pipeline. In this case also, existing fragments can be reused as is and the rendering step to operationalise the image processing pipeline under consideration performed on the current semantic community memory content. If no high-level end-user's intentional requirements are already available, the domain expert specifies the image processing pipeline under consideration with the help of the community semantic memory manager.

Then, for each subsection identified in the high-level abstract fragment, the domain expert may search for existing fragments supporting their operationalisation. If it is the case, then s/he can decide to rely on them and stop the authoring process. Otherwise, s/he may prefer to provide his/her own way to operationalise the sub-goals. By doing so, the domain expert enriches the semantic community memory with alternative ways to operationalise already registered goals. This will result in enriching the operationalisation means of the image processing pipelines already formalised into fragments stored in the semantic community repository. In fact, when someone else looking for the sub-goals under consideration will perform a rendering process, if his/her image processing pipeline relies on the achievement of a target intention for which a new operationalisation means has been provided, previously stored in the semantic community repository as well as the new ones are exploited, increasing the number of ways to find suitable Web Service descriptions. Each time the domain expert, with the help of the community semantic memory manager, decides to provide new ways to operationalise a map section, s/he has to select the right level of specification of the fragment signature, in order to allow the reuse of the fragment under construction outside of the scope of the image processing pipeline under consideration.

From a more general point of view, domain and community semantic memory managers mainly provide fragments: Domain experts focus on high-level fragments, close to the image processing pipelines they want to operationalise. community semantic memory managers focus on low level fragments, that is to say fragments operationalising basic image processing pipelines. And service designers mainly focus on providing fragments to promote existing Web Services. But domain and community semantic memory managers may also provide fragments to specify their requirements in term of services. And the service designers may also provide fragments in order to show examples of use of available Web Services inside the scope of more complex examples of image processing pipelines. By relying on a rule-based specification to operationalise reusable fragments and by providing distinct and dedicated modeling techniques to both service designers and service final-users as well as mapping mechanisms between them, we assist the bidirectional collaboration between neuroscientists and computer scientists inside the community.

An important objective of the SATIS project is to provide to domain experts means to better understand what are the characteristics of the available services and how to use them in the scope of the image processing pipeline they are interested in. We support this aim by several means. The SATIS approach relies on a controlled vocabulary (domain ontology) to qualify Web Services as well as requirements, this way reducing the diversity in the labelling, especially in Web Services descriptions elements. Requirements about Web Services are described in terms of intentions and strategies that is to say a vocabulary familiar to the domain expert, making the understanding of the a Web Service purpose easier to understand by domain experts. We propose to specify required Web Service functionalities in terms of queries (*i.e.* generic Web Service descriptions) instead of traditional Web Service descriptions in order to provide an abstraction level supporting the categorisation of available Web Services and this way an easier understanding of the content of the registry by domain experts.

In our approach we clearly distinguish an authoring step and a rendering step. During the authoring step, the focus is on the elicitation of the search procedure. The domain experts think in terms of intentions and strategies (and not in terms of services). His/her search procedure is fully described, eventually with the help of the fragments already present in the community semantic memory. During the rendering step, it is the system (and not the domain expert) which tries to find Web Services corresponding to the requirements specified by the experts (by proving goals and sub-goals). Indeed, the experts don't need at all to know the content of the registry. A pertinent subset of it will be extracted by the system and shown to the experts.

And finally, SATIS relies on a fragment based approach which doesn't show to the domain expert the full set of rules exploited by the backward chaining engine to satisfy the user requirements. When rendering a search procedure, the domain expert only selects the intention characterizing his/her image processing pipeline and the system will search for the rules to use. A set of Web Services descriptions is given to the domain expert as result. But the complexity and the number of rules used to get the solution are hidden to the domain expert.

### 3.2 SATIS Architecture

In Illustration 1, we describe entities (classes) and relationships between these entities in SATIS. We can see traditional entities of the map model (to express and capture goals) : *Map*, *Section*, *Strategy* and *Goal*.

In Illustration 2, we can see, at the top, an example of a map which is a graphical way to express and capture goals of end-users. Sections are parts of map. On this map, we have two main goals (*Make an image homogeneous* and *Put image in a Single reference* ; *Start* and *Stop* are mandatory special goals) and five main strategies (*by normalization*, *by debiasing*, *by denoising*, *by registration* and *by rotation* ; the arrow between *Put image in a Single reference* and *Stop* is another, anonymous, strategy). The Illustration 2 will be presented more precisely in the next section.

In Illustration 1, some entities are more specific to SATIS, and are made in order to capitalise approaches (*Fragment*, *Directive* and *Intentional Directive*) and to search adequate Web Services (*Operational Directive* and *SPARQL Request*).

The main SATIS process consists to find semantic annotation of Web Services from end-users' intentional goals. This process is constituted by a set of relatively autonomous fragments which are expressed at different granularity level. Thus, a fragment is an autonomous and coherent piece of the process of the Web Services research. This is a modular point of view in order to facilitate the adaptations and extensions of the approach. Moreover, this modularity allows to reuse fragments which have been previously designed to implement another approach.

The body of a fragment is constituted by directives which are autonomous and reusable. The fragment signature gives the adequate situation (context) to reuse this fragment. A directive is a fact, an indication or a procedure to determine the way to realise an action. For SATIS, a directive is more precisely a know-how about the mean to reach a goal from a given situation. We distinguish two kinds of directive: Intentional directives and operational directives. An intentional directive is used to specify an high-level intentional need. This kind of need has to be refined in more precise ones. An operational directive represents some generic descriptions of Web Services.

To conclude this short presentation of the SATIS core, let's define intentional and operational fragments. Intentional ones are dedicated to search adequate Web Services, they are constituted by a section which gives the source situation and a request (in SPARQL or another request language) to execute the research. An operational fragment needs to refinements to be concrete: it is constituted by a section which gives the source situation and a map to refine.

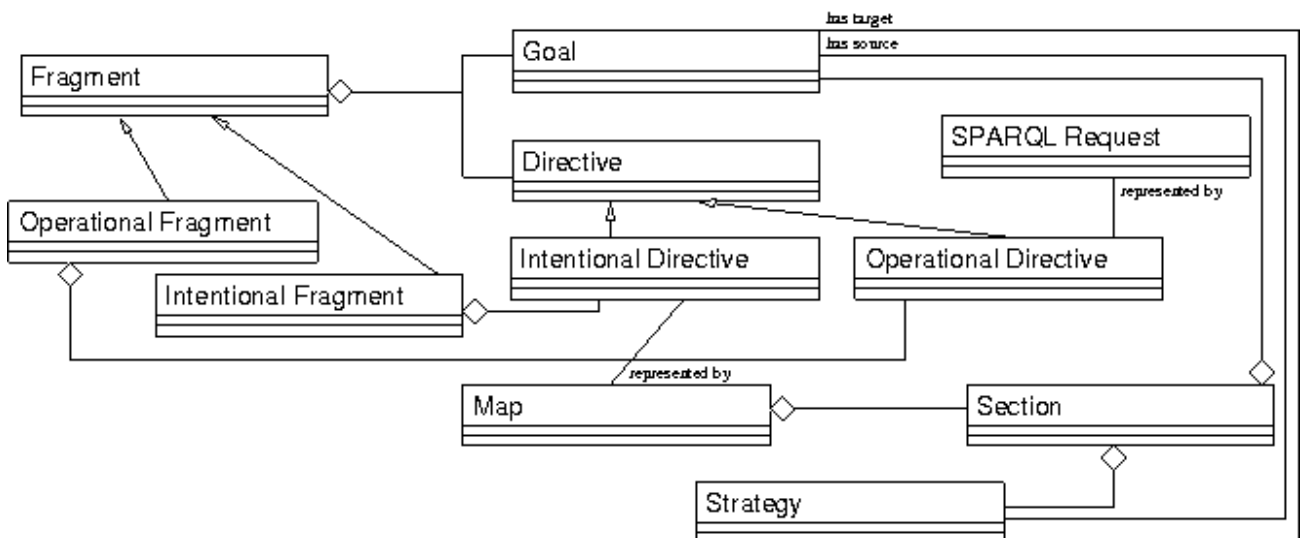

*Illustration 1: Fragment Model*

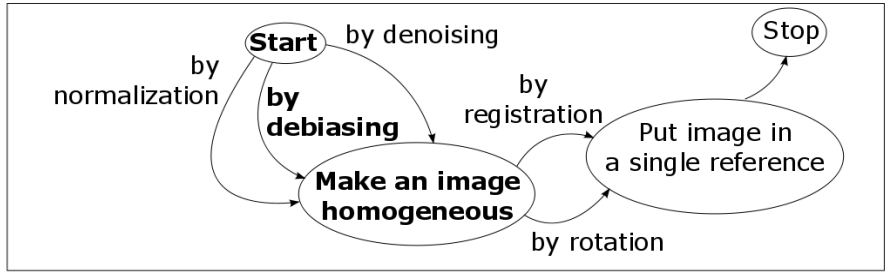
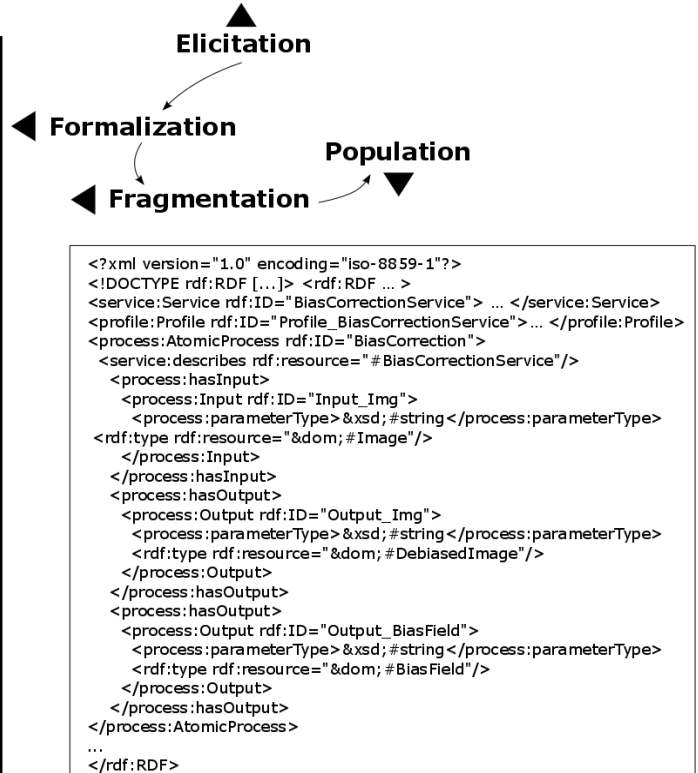

*Illustration 2: SATIS*

# 4 Semantic Annotation to Support Workflows Combination Know-How Capitalisation

In our approach, we adopt Web semantic languages and models as a unified framework to deal with (i) high-level end-user's intentional requirements, (ii) generic Web Service descriptions and (iii) Web Service descriptions themselves. With regards to high-level end-user's intentional requirements, we adapted the map model [2] to our concern and gathered its concepts and relationships into an RDFS [4] ontology dedicated to the representation of intentional processes: the map ontology [36]. As a result, image processing pipelines annotated with concepts and relationships from this ontology can be shared and exploited by reasoning on their representations. Semantic *Web Service descriptions* are specified with the help of the OWL-S ontology [5] as well as a domain ontology. And finally, *generic Web Service descriptions* are specified with the help of the W3C standard query language for RDF [3] annotations: SPARQL [37]. Indeed, generic Web Service descriptions are formalised into graph patterns over Web Services descriptions.

## 4.1 Formalisation of high-level end-user's intentional requirements

To further formalise map elements, we rely on [38] proposal, which has already proved to be useful to formalise goals [39][40][2]. According to [38], an intention statement is characterised by a verb and some parameters which play specific roles with respect to the verb. Among the parameters, there is the object on which the action described by the verb is processed. We gathered the concepts and relationships of the map model and this further formalisation into an RDFS [4] ontology dedicated to the representation of intentional processes: the map ontology [36].

Based on the map ontology, image processing pipelines (or fragments of image processing pipeline) are then represented by RDF annotations.

The mappings between the domain ontology and the map ontology are automatically created when concepts of the domain ontology are selected to formalise map content. Domain concepts are then considered as instances of *AnyVerb*, subclass of *Verb*, *AnyObject*, instance of *Object* and *AnyParameter,* instance of *Parameter*. Verb, Object and Parameter are provided by the map ontology while AnyVerb, AnyObject and AnyParameter are provided in the mappings between the domain ontology and the map ontology to support reasoning.

Let us consider again our running example and the map depicted in the upper part of the Illustration 2. The highlighted section has a start intention as source intention, a target intention *labelled "Make an image homogeneous" and a strategy* labelled "by debiasing". Thanks to the domain and the map ontologies as well as the mappings between them, the target intention is described by its verb *Homogenise* and its object *Image* ; the strategy is described by its parameter *Debiasing* and the source intention is described by its verb *AnyVerb* and its object *AnyObject*. The RDF dataset shown in the left part of Illustration 2 where namespace *map* refers to the map ontology and namespace *dom* refers to a domain ontology corresponds to the formalisation of the highlighted section in the map under consideration.

By relying on RDF(S) which is now a widespread Web standard, we ensure the capitalization, reuse and share of these representations of search procedures among community members. Beyond an alternative way to organize and to dynamically access resources in a community memory, we provide means to capitalize search procedures themselves. We take advantage of the inference capabilities provided by the RDF framework to reason on search process representations, especially to organize them and retrieve them for reuse.

### 4.2 Formalisation of generic Web Service descriptions

In SATIS, we assume Web Service descriptions are expressed in OWL-S. In our current scenarios, we only use the profile and the grounding of OWL-S as well as the input and output specifications in the process description. We enrich OWL-S description by considering their content (as input and output parameters for example) as instances of domain concepts. Thanks to this additional instantiation of domain concepts, it makes it possible to reason on OWL-S description element types to retrieve for instance subclasses of concepts we are interested in. An example of excerpt of such an enriched OWL-S Web Service description is shown on the right side of Illustration 2. This description deals with a Web Service requiring an image as input and providing a debiased image and a bias field as output.

As we don't want community members to strongly couple high-level intentional requirements to technical Web Service specifications, we introduced generic Web Service descriptions aiming at qualifying the required features when looking for Web Services to operationalise image processing pipelines (or fragments of image processing pipeline). For instance, by looking for a Web Service which takes as input an image and provides as output a debiased image, the end-user specifies the kind of Web Service s/he is interested in without explicitly referring to one specific Web Service. The Web Service which description is shown in Illustration 2 will be retrieved by rendering the query shown on the left bottom side of Illustration 2. But if other Web Services also deal with image (or a subclass) and debiased image (or a subclass), they will also be retrieved by the query under consideration. By doing so, we assume a loosely coupling between high-level end-user's intentional requirements on one hand and Web Services descriptions on the other hand: if new Web Service descriptions are added inside the community semantic memory, they can be retrieved to operationalise a high-level end-user's intentional requirement even if the requirement has been specified before the availability of the Web Services under consideration; and if Web Service descriptions are removed from the community semantic memory, the high-level end-user's intentional requirements that they satisfied are still valid and may be operationalised by other available Web Services. Web Services are dynamically selected when rendering queries associated to high-level end-user's intentional requirements.

Generic Web Service descriptions are expressed as SPARQL queries among the Web Service descriptions expressed in OWL-S. An example of such a generic Web Service description is shown in the left bottom side of Illustration 2. This query aims at searching for Web Service descriptions having as input an *Image* (or a subclass of it) and as output a *DebiasedImage* (or a subclass of it) and a *BiasField* (or a subclass of it).

### 4.3 Rules

In SATIS, the process consisting in retrieving Web Services descriptions from high-level end-user's intentional requirements about image processing pipelines is viewed as a set of loosely coupled fragments expressed at different levels of granularity. A fragment is an autonomous and coherent part of a search process supporting the operationalisation of part of an image processing pipeline by Web Services. Such a modular view of the process aiming at retrieving Web Service descriptions from high-level end-user's intentional requirements favours the reuse of fragments authored to deal with a specific high-level end-user's image processing pipeline in the building of other pipelines.

The fragment body captures guidelines that can be considered as autonomous and reusable. The fragment signature captures the reuse context in which the fragment can be applied.

For us, a guideline embodies know-how about how to achieve an intention in a given situation. We distinguish two types of guidelines: intentional and operational guidelines. Intentional guidelines capture high-level end-user's intentional requirements which have to be refined into more specific requirements. Operational guidelines capture generic Web Service description.

Map formalisations and SPARQL queries respectively constitute the body of intentional and operational reusable fragments. The fragment signature characterises the fragment content and let the other members of the community understand in which situation the fragment may be useful. A fragment signature aims at capturing the purpose of the fragment and its reuse situation. A fragment signature is specified by a map section. The target intention of the section indicates the goal of the reusable fragment and the source intention as well as the strategy specify the reuse situation in which the fragment is suitable.

In a fragment signature, the target intention is mandatory to elicit the goal of the fragment. If a particular context is required to use the fragment, a source intention or a strategy may be used to specify it. Web Services retrieved by rendering a fragment which signature does not include source intention are less precise than Web Services retrieved by rendering a fragment which signature is fully described by a source intention in addition to the target intention. Similarly, to specify or not a strategy can respectively reduce or enlarge the spectrum of Web Services considered. Both means (specifying a source intention or a strategy) can be combined to obtain a signature which genericity level actually corresponds to the guidelines proposed in the body of the fragment.

The RDF dataset shown on the left side of Illustration 2 corresponds to a fragment signature which body is the SPARQL query shown on the left bottom side of Illustration 2. The black frame depicts a fragment. As its body contains a query, it is an operational fragment. An example of intentional fragment may be a fragment to perform image preprocessing body provides a RDF dataset formalizing the full map (i.e. all the sections) shown on top of Illustration 2 and is specified by a section which target intention is finalised by the object *Image* and the verb *Preprocessing*.

This intentional fragment capitalize a know-how about how to break down an image preprocessing high-level requirement into sub-goals in order to find Web Services to operationalise an image processing pipeline.

Indeed in SATIS, fragments are implemented by backward chaining rules, which conclusions represent signatures of fragments and which premises represent bodies of fragments (either operational or intentional guidelines). We call a rule *concrete* or *abstract* depending on whether its premise encapsulates operational or intentional guidelines.

These rules are implemented as SPARQL construct queries. The CONSTRUCT part is interpreted as the head of the rule, the consequent that is proved. The WHERE part is interpreted as the body, the condition that makes the head proved. When considered recursively, a set of SPARQL construct queries can be seen as a set of rules processed in backward chaining.

# 5  Conclusion

The ability to support adequacy between service users needs and service providers proposals is a critical factor for achieving interoperability in distributed applications in heterogeneous environments such as scientific workflows. The problem of discovery, or how to retrieve existing workflows that can be reused or re-purposed for one's goals, is especially important for end users who have little knowledge about workflows.

In this chapter, we focused on knowledge engineering to support service combination from an end-user perspective. We proposed SATIS, a framework to turn this know-how into reusable guidelines or best practices and to provide means to support its capitalization, dissemination and management inside business users communities.

SATIS offers the capability to capture high-level end-user's requirements in an iterative and incremental way and to turn them into queries to retrieve Web Services descriptions. The whole framework relies in reusable and combinable elements which can be shared out inside the scope of a community of users. In our approach, we adopt Web semantic languages and models as a unified framework to deal with (i) high-level end-user's intentional requirements, (ii) generic Web Service descriptions and (iii) Web Service descriptions themselves. SATIS aims at supporting collaboration among the members of a neuroscience community by contributing to both mutualisation of high-level intentional specification and cross-fertilisation of know-how about Web Services search procedures among the community members.

Future works will focus on enriching the formalisation step by taking into account additional information in order, for instance, to derive criteria related to quality of services. Indeed, we plan to extend our Web Service annotation model with quality of service (QoS) information and to qualify map strategies by QoS domain concepts. We will also concentrate on providing query patterns to help experts writing generic Web Service descriptions.